\documentclass[preprint]{revtex4-1}
\usepackage{pifont}
\usepackage{graphicx,colordvi}
\usepackage{amsmath,amssymb,amsthm}
\usepackage[colorlinks=true,urlcolor=blue,linkcolor=blue,citecolor=red]{hyperref}
\usepackage{mathrsfs}

\newcommand{\R}{\mathbb{R}}
\newcommand{\ini}{\begin{equation}}
\newcommand{\fin}{\end{equation}}
\newcommand{\inia}{\begin{eqnarray}}
\newcommand{\fina}{\end{eqnarray}}

\newcommand {\begproof} {\noindent {\bf Proof.} }

\newtheorem{lemma}{Lemma}

\newtheorem{proposition}{Proposition}
\newtheorem{corollary}{Corollary}

\begin{document}

%\draft
\title{\bf  Qualitative Aspects of the Solutions of a Mathematical Model
for the Dynamic Analysis of the Reversible Chemical Reaction
$\mbox{SO}_{2\,(\mathrm{g})}\,+\,
\frac{1}{2}\,\mbox{O}_{2\,(\mathrm{g})}\,\rightleftharpoons \,
\mbox{SO}_{3\,(\mathrm{g})}$ in a Catalytic Reactor}
\author{$^{1}$Angulo Wilfredo  and $^{2}$Contreras Joyne}
\address{$^{1}$Decanato de Ciencias y Tecnolog\'{\i}a, Universidad Centroccidental Lisandro Alvarado, Barquisimeto,Venezuela\\
$^{2}$Decanato de Agronom\'{\i}a, Universidad Centroccidental Lisandro Alvarado, Barquisimeto,Venezuela}

\begin{abstract}
We present some qualitative aspects concerning the solution to the
mathematical model describing the dynamical behavior of the
reversible chemical reaction $\mbox{SO}_{2\,(\mathrm{g})}\,+\,
\frac{1}{2}\,\mbox{O}_{2\,(\mathrm{g})}\,\rightleftharpoons \,
\mbox{SO}_{3\,(\mathrm{g})}$ carried out in a catalytic reactor used
in the process of sulfuric acid production.

%The electrostatic potential is calculated and $1/r^{2\kappa}$, with $\kappa$ the vector coupling parameter, is estimated as the order of corrections generated by the massive gauge fields set.

\vspace{0.5 cm}
PACS numbers: %04.20.-q, 11.27.+d, 04.50.+h
\end{abstract}

\maketitle

\section{Introduction}\label{section1}

The production of most industrially important chemicals involves
catalysis. Catalysis is relevant to many aspects of environmental
science, e.g. the catalytic converter in automobiles and the
dynamics of the ozone hole. Catalytic reactions are preferred in
environmentally friendly green chemistry due to the reduced amount
of waste generated, as opposed to stoichiometric reactions in which
all reactants are consumed and more collaterals products are formed.
Particularly, the oxidation of sulfur dioxide to sulfur trioxide
using oxygen or air and a suitable catalyst such as vanadium
pentoxide is well known for the sulfuric acid production. In this
sense, the idea that the performance of these continuous catalytic
processes under invariable conditions is highly efficient has gained
great popularity, among chemical engineers that design catalytic
reactors where the reaction will be carried out \cite{ca}. However,
very often the optimal conditions of the process can be achieved
with the unsteady-state operation and the steady-state operation
will be a particular case of the unsteady-state conditions.
Unsteady-state operation broadens the possibilities to form the
profiles of the catalyst states, concentrations, and temperatures in
reactors, thus providing more favorable conditions for the process
performance \cite{mm}. Research work like this involve many areas
of chemistry and physical-chemistry, but mathematical modeling
is an important tool for rapid and reliable reactor development and
design \cite{carberry}. The models are built from the basic studies
of the reaction mechanism and kinetics, the transfer processes, and
the interactions within the system. A detailed understanding of the
elementary processes enables the construction of powerful and
complex models for dynamic and steady-state simulation. With the aid
of experimentally determined parameter values we can develop new
processes or improve existing ones using dynamical simulations
based on its mathematical models \cite{bales}.

In this work, we present a mathematical model for the dynamical
analysis of the reversible chemical reaction associated to the
oxidation of sulfur dioxide to sulfur trioxide using oxygen in
presence of the vanadium pentoxide catalyst, and we study some
qualitative aspects concerning its solution as a previous step for
the simulation of the catalytic reactor where the reaction will be
carried out.

This paper is organized as follows. In Section \ref{section2} we
present the mathematical model formulated as a problem of Cauchy
or initial conditions for the state variables that they define to
the studied catalytic system, using as reference a model presented
in \cite{ctacv}. In Section \ref{section3} we begin the qualitative
study of the mathematical model demonstrating that this is a
well-posed problem; in addition we present the characteristics of
the set of steady-state. Next, Section \ref{section4} is devoted to
the study the solutions of the dynamics states for the system, we present
the qualitative aspects concerning the behavior when the operation time is very long. In Section
\ref{section5} we present a brief discussion from the
physicochemical point of view and we finalize with the conclusions
of this research in Section \ref{section6}.

\section{Mathematical model}\label{section2}

\subsection{Description of the catalytic system}
The studied catalytic system was the oxidation of sulfur dioxide
(SO$_{2}$) to sulfur trioxide (SO$_{3}$) in presence of the vanadium
pentoxide catalyst (Vn$_{2}$O$_{5}$). The stoichiometric equation is:
\begin{equation}\label{eq1}
  \mbox{SO}_{2\,(\mbox{g})}\,+\, \frac{1}{2}\,\mbox{O}_{2\,(\mbox{g})}\,\rightleftharpoons \,
  \mbox{SO}_{3\,(\mbox{g})}.
\end{equation}
This reaction is exothermic in the forward direction, denoted by $\rightharpoonup$, and endothermic in the reverse direction, denoted by $\leftharpoondown$. Also, the reaction is a homogenous mixture, its reactans and products are in gaseous phase relative to the conditions of operation in the bed of the catalytic reactor. The speed of this reaction has been widely studied and the expression that suits best is the Eklund's equation
( see \cite{mm}):
\begin{equation}\label{eq2}
  r_{\mbox{SO}_{2}}=k\sqrt{\frac{p_{\mbox{SO}_{2}}}{p_{\mbox{SO}_{3}}}}
  \left[p_{\mbox{O}_{2}}-\left(\frac{p_{\mbox{SO}_{3}}}{p_{\mbox{SO}_{2}}k_{p}}\right)^{2}\right],
\end{equation}
where $r_{\mbox{SO}_{2}}$ is the reaction rate referred to the
SO$_{2}$ ($\mbox{mol SO}_{2}/\mbox{s$\cdot$gr of the catalyst}$),
$p_{i}$ is the partial pressure (atm) of the $i$-\emph{th} component
($i=\mbox{SO}_{2},\mbox{ SO}_{3},\mbox{ O}_{2}$), $k$ is the kinetic
coefficient of reaction rate and $k_{p}$  the coefficient of
chemical equilibrium, both as a function of the temperature (for
more details, see \cite{ca} and \cite{fo}).

\subsection{Formulation of the mathematical model}

For the sake of simplicity, we consider a fixed volume element of
catalyst bed, with cylindrical geometry of finite length $L$ and
radius $R$, in which the reaction is carried out. We assume that
gradients of concentration and temperature in the radial direction
($0\leq r\leq R$) of the catalyst bed do not exist. These gradients
are more noticeable in the longitudinal direction ($0\leq z\leq L$),
but this spatial variation is not considered for the dynamic study
that we will address in this work. Finally, we consider only the
variation of the concentration and temperature with respect to the
time, and we assume that the changes in the total pressure of the
system with respect to time are negligibles at each bed's output, therefore a balance of momentum was not needed.

The complete problem of interest, obtained by a dynamic balance of matter and
caloric energy, is described by the following equations:

\begin{eqnarray}
\frac{dX_{\mbox{A}}(t)}{dt}&=& \frac{-r_{\mbox{A}}\left(1+\epsilon
X_{\mbox{A}}(t)\right)\left(1-\phi\right)\rho
_{c}}{C_{\mbox{A}_{0}}},\nonumber\\[-1.5ex]
\label{eq3} \\[-1.5ex]
\frac{dT(t)}{dt}&=&\frac{-r_{\mbox{A}}\left(1+\epsilon
X_{\mbox{A}}(t)\right)\left(1-\phi\right)\rho _{c}\left(-\Delta
Hr\right)}{C_{\mbox{A}_{0}}\left(\sum_{i}\theta _{i}Cp_{i}+
X_{\mbox{A}}(t) \Delta
Cp\right)+C_{\mbox{I}}Cp_{\mbox{I}}},\nonumber
\end{eqnarray}
with the initial data $X_{\mbox{A}}(0)=X_{\mbox{A}_{0}}$ and $T(0)=T_{0}$ for
the state variables $X_{A}$ and $T$ respectively. The subscript ($\mbox{A}$)
was used to denote component SO$_{2}$ and subscript ($\mbox{I}$) to
denote inert present in the mixture such as bimolecular nitrogen. Therefore, $X_{\mbox{A}}$
represents the molar conversion of the $\mathrm{SO}_2$ in the mixture,
$r_{\mbox{A}}=f(X_{\mbox{A}},T)$ is the Eklund's expression written in
terms of the molar conversion and of the temperature $T$ of the system.
On the other hand $\epsilon$, $\phi$, $\rho_{c}$, $C_{\mbox{A}_{0}}$, $Cp_{i}$, $\theta_{i}$, $\Delta Cp$,
$C_{\mbox{I}}$ and $Cp_{\mbox{I}}$ are (constant) given physical parameters.
This model is complemented with the following relations:
\begin{itemize}
\item Coefficient of chemical equilibrium
\begin{equation}\label{eq5}
k_{p}=\exp{\left(\frac{11829\text{.}44}{T}-11\text{.}24\right)}.
\end{equation}
\item Kinetic coefficient of reaction rate
\begin{equation}\label{eq6}
k=\exp{\left(\frac{-97782\text{.}22}{T}-110\ln{(1\text{.}8T)}+912\text{.}8\right)}
\end{equation}
\item Heat of reaction
\begin{equation}\label{eq7}
\Delta H_{r}=34923\text{.}286-65\text{.}395T+0\text{.}0725T^{2}.
\end{equation}
\end{itemize}

\subsection{Abstraction of the mathematical model}

We begin redefining the two state variables
(conversion of the SO$_{2}$ and temperature of the system) as follows:
\[ u_{1}=u_{1}(t)=X_{\mbox{A}}(t)\,\,\mbox{ and
}\,\,u_{2}=u_{2}(t)=T(t),
\]
such that, for all time $t\in[0,+\infty)$,
$\mathbf{u}(t)=\left(u_{1}(t),u_{2}(t)\right)$ is the vectorial function
of the two state variables to determine in the subset $\bar{\Omega}\subset\R^2$ given by
\[
\bar{\Omega}=[0,u_{1}^{\ast}]\times[u_{2}^{-},u_{2}^{+}],
\]
with $u_{1}^{\ast}=1-\delta$ for $0< \delta <1$, and where
$u_{2}^{-}$ and $u_{2}^{+}$, respectively, are taken as
\begin{equation}\label{eq8}
 u_{2}^{-}=\min\limits_{t\in[0,+\infty)}\{0<T(t)<\infty\}\hspace{0.3cm}\mbox{and}
 \hspace{0.3cm}u_{2}^{+}=\max\limits_{t\in[0,+\infty)}\{0<T(t)<\infty\}.
\end{equation}
The right side of each EDO in (\ref{eq3}) is a real valued function defined on $\bar{\Omega}$:
\[
f_{1}(\mathbf{u})=\beta_{1}r_{A}(\mathbf{u})g_{1}(u_{1})\hspace{0.3cm}\mbox{and}\hspace{0.3cm}
f_{2}(\mathbf{u})=f_{1}(\mathbf{u})g_{2}(\mathbf{u}),\]where
\[
g_{1}(u_{1})=1+\epsilon u_{1},\hspace{0.3cm}
g_{2}(\mathbf{u})=\frac{b_{1}+b_{2}u_{2}+b_{3}u_{2}^{2}}{\beta_{2}+\beta_{3}u_{1}}\hspace{0.3cm}\mbox{and}\hspace{0.3cm}
r_{A}(\mathbf{u})=K_{1}(u_{2})K_{2}(u_{1})K_{3}(\mathbf{u}),
\]
with
\[
K_{1}(u_{2})=\exp{\left(\frac{a_{1}}{u_{2}}+a_{2}\ln{(a_{3}u_{2})}+a_{4}\right)},\hspace{0.3cm}
K_{2}(u_{1})=\sqrt{\frac{1-u_{1}}{\theta_{c}+u_{1}}},
\]
\[K_{3}(\mathbf{u})=L_{1}(u_{1})-L_{2}(u_{1})L_{3}(u_{2}).\]
For $K_{3}$, functions $L_{1}$, $L_{2}$ and $L_{3}$ are given by:
\[
L_{1}(u_{1})=\frac{a_{5}-a_{6}u_{1}}{g_{1}(u_{1})},\hspace{0.3cm}
L_{2}(u_{1})=\left(\frac{\theta_{c}+u_{1}}{1-u_{1}}\right)^{2}\hspace{0.3cm}\mbox{and}\hspace{0.3cm}
L_{3}(u_{2})=\frac{1}{\left(\exp{\left(\frac{a_{7}}{u_{2}}+a_{8}\right)}\right)^2}.
\]
Here, $\beta_{1}$, $\beta_2$, $\beta_3$, $\theta_{c}$, $a_3$, $a_4$,
$a_5$, $a_6$, $a_7$, $b_1$ and $b_3$ are constant \emph{strictly
positive}; $\epsilon$, $a_1$, $a_2$, $a_8$, $b_2$ and $b_4$ are
constant \emph{strictly negative}. For some of these constants, the
physicochemical behavior of the system provides the following
restrictions:
\begin{eqnarray}
\beta_{2}&>&\beta_{3}>\beta_{1},\nonumber\\[-1.5ex]
\nonumber\\[-1.5ex]
 0&<&\theta_{c}\ll 1,\nonumber\\[-1.5ex]
 \nonumber\\[-1.5ex]
a_{7}>a_{4}&>&a_{3}>a_{5}>a_{6},\nonumber\\[-1.5ex]
\nonumber\\[-1.5ex]
a_{8}&>&a_{2}>a_{1},\nonumber\\[-1.5ex]
\nonumber\\[-1.5ex]
b_{1}&>&b_{3},\nonumber\\[-1.5ex]
\nonumber\\[-1.5ex]
\left|\epsilon\right|&<&1,\nonumber\\[-1.5ex]
\nonumber\\[-1.5ex]
-1/\epsilon&\gg& 1,\nonumber\\[-1.5ex]
\nonumber\\[-1.5ex]
b_{2}^{2}&\ll& 4b_{3}b_{1}.\nonumber
\end{eqnarray}

With all the above, the functions $f_{1}$  and $f_{2}$ define the components of a vectorial field (of directions):
\begin{equation}\label{eq9}
\mathbf{f}:\bar{\Omega}\subseteq\mathbb{R}^2\rightarrow\mathbb{R}^2
\end{equation}and the mathematic model (\ref{eq3})is rewritten as the problem of Cauchy,
or initial conditions, for two nonlinear
ordinary differential equations: given the
vector $\mathbf{u}^{0}\in\bar{\Omega}$, to find
$\mathbf{u}\in\bar{\Omega}$ solution of
\begin{eqnarray}
\frac{d\mathbf{u}(t)}{dt}&=&\mathbf{f}(\mathbf{u}(t)),\hspace{0.3cm}\forall
t\in[0,+\infty),\nonumber\\[-1.5ex]
\label{eq9.1} \\[-1.5ex]
\mathbf{u}(0)&=&\mathbf{u}^{0}.\nonumber
\end{eqnarray}

\section{Solutions of the mathematical model}\label{section3}

\subsection{Solutions of steady-state}

The dynamic analysis of a chemical reaction by means of a
mathematical model begins by the determination of the stationary states. For the reaction studied in this paper the steady-states
are given by the following subset:
$$\Gamma=\left\{\mathbf{u}^{e}=(u^{e}_{1},u^{e}_{2})\in\bar{\Omega};\hspace{0.2cm}\mathbf{f}(\mathbf{u}^{e})
=\mathbf{0},\hspace{0.1cm}u_{1}^{e}\in
[0,u_{1}^{\ast}],\hspace{0.1cm}\mbox{whit}\hspace{0.2cm}u_{1}^{e}\neq-\frac{1}{\epsilon},u_{2}^{e}=h(u_{1}^{e})\right\},$$
where $h:[0,u_{1}^{\ast}]\rightarrow\R$ is defined as
$$h(u_{1})=\frac{a_{7}}{\ln{\biggl(\sqrt{\frac{L_{2}(u_{1})}{L_{1}(u_{1})}}\biggr)}-a_{8}},$$for
which it is easily verifiably that $\lim\limits_{u_{1}\rightarrow
1}{h(u_{1})=0}$ and thus, $u_{2}^{-}=h(u_{1}^{\ast})$ and $u_{2}^{+}=h(0)$.

\begin{figure}[!htb]
\centering
\includegraphics[width=6.5cm, height=4.5cm,keepaspectratio]{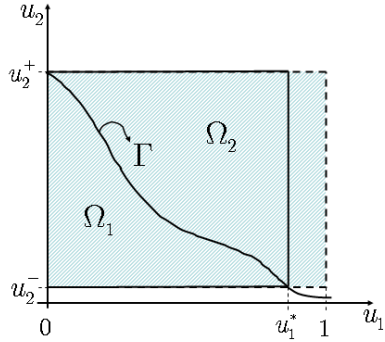}
\caption{Continuous of steady-states $\Gamma$ and subsets
$\Omega_{1}$ and $\Omega_{2}$}\label{geometria1}
\end{figure}

The subset previously defined divides the set $\Omega$ into two
simply connected subdomains $\Omega_{1}$ and $\Omega_{2}$:
\begin{eqnarray}
\Omega_{1}&=&\{\mathbf{u}=(u_{1},u_{2});\hspace{0.2cm}u_{2}^{-}<u_{2}<u_{2}^{e},
\hspace{0.1cm}\forall u_{1}\in(0,u_{1}^{\ast})\},\nonumber\\[-1.5ex]
\nonumber\\[-1.5ex]
\Omega_{2}&=&\{\mathbf{u}=(u_{1},u_{2});\hspace{0.2cm}u_{2}^{e}<u_{2}<u_{2}^{+},
\hspace{0.1cm}\forall u_{1}\in(0,u_{1}^{\ast})\};\nonumber
\end{eqnarray}
indeed $\Gamma=\bar{\Omega}_{1}\cap\bar{\Omega}_{2}$. Figure
\ref{geometria1} illustrates the continuous $\Gamma$ of
steady-states and the subsets $\Omega_{1}$ and $\Omega_{2}$.

\subsection{Existence and uniqueness of the solutions of dynamic state}

The global existence and uniqueness of the solutions of dynamic
state for the problem (\ref{eq9.1}), are a direct consequence of the
associated global Lipschitz property to the vectorial field
$\mathbf{f}$ on $\Omega\cup\Gamma$. Simultaneously, this property is
a direct consequence of the existence and boundedness of the
partials drivative $\partial f_{i}/\partial u_{j}$, $i,j=1,2$, on
$\Omega\cup\Gamma$ (see \cite{ti}, \cite{cl}, \cite{ah} and
\cite{ii}), associated to the component functions (scalar fields)
$f_{i}$. This is the objective of the following proposition.
\begin{proposition}\label{propo1}
For each scalar field $f_{i}:\Omega\cup\Gamma\rightarrow\R$,
$i=1,2$, the partial derivative $\partial f_{i}/\partial u_{j}$,
$i,j=1,2$, exist and is bounded on $\Omega\cup\Gamma$.
\end{proposition}
\begproof
Thanks to the structure of each scalar field $f_{i}$ there exists
$\partial f_{i}/\partial u_{j}$ for $i,j=1,2$. Indeed, we have:
\begin{eqnarray}
\frac{\partial f_{1}}{\partial
u_{1}}&=&\beta_{1}\left[g_{1}\frac{\partial
r_{A}}{\partial u_{1}}+\epsilon r_{A}\right],\nonumber\\[-1.5ex]
\label{eq18}\\[-1.5ex]
\frac{\partial f_{1}}{\partial u_{2}}&=&\beta_{1}g_{1}\frac{\partial
r_{A}}{\partial u_{2}};\nonumber
\end{eqnarray}
\begin{eqnarray}
\frac{\partial f_{2}}{\partial u_{1}}&=&g_{2} \frac{\partial
f_{1}}{\partial u_{1}}+f_{1}
\frac{\partial g_{2}}{\partial u_{1}},\nonumber\\[-1.5ex]
\label{eq19}\\[-1.5ex]
\frac{\partial f_{2}}{\partial u_{2}}&=&g_{2}\frac{\partial
f_{1}}{\partial u_{2}}+f_{1}\frac{\partial g_{2}}{\partial
u_{2}}.\nonumber
\end{eqnarray}
Finally, as all the previous derivative are functions composed of
continuous and bounded elementary functions on $\Omega\cup\Gamma$,
then these derivatives also are continuous and bounded functions on
$\Omega\cup\Gamma$ \cite{pz, rb}, i.e, for all well-know and fixed physicochemical parameters, there exists a constant
$C^{i}_{j}$, for $i,j=1,2$, depending only on $\delta$
such that\[\left|\frac{\partial f_{i}}{\partial u_{j}}\right|<
C_{j}^{i}.\]Indeed, taking the absolute-value in both members from
the expressions (\ref{eq18}) and (\ref{eq19}) we have:
\[
C^{1}_{1}=\frac{\beta_{1}M_{1}}{\theta^{^{1/2}}}
\left[\frac{\Lambda_{1}(\theta_{c},\delta)\Lambda_{2}(\theta_{c},\delta)}
{\biggl(\exp\biggl(\frac{a_{7}}{u_{2}^{+}}+a_{8}\biggr)\biggr)^{2}}
+\frac{1}{|1+\epsilon(1-\delta)|}\biggl((\Lambda_{3}(\theta_{c},\delta)+1)a_{5}
+\frac{a_{6}+a_{5}}{|1+\epsilon(1-\delta)|}\biggr)\right],\]

\[C_{2}^{1}=\frac{\beta_{1}M_{1}}{\theta_{c}^{^{1/2}}(u_{2}^{-})^{2}}
\left[\frac{\delta^{2}\Lambda_{1}^{2}(\theta_{c},\delta)(|a_{2}|u_{2}^{-}+a_{1}+2a_{7})}
{\biggl(\exp\biggl(\frac{a_{7}}{u_{2}^{+}}+a_{8}\biggr)\biggr)^{2}}
+\frac{a_{5}(|a_{2}|u_{2}^{-}+a_{1})}{|1+\epsilon(1-\delta)|}\right];\hspace{6cm}\]

\[C_{1}^{2}=\biggl(\frac{b_{1}+|b_{2}|u_{2}^{-}+b_{3}(u_{2}^{+})^{2}}
{\beta_{2}}\biggr)\left[C_{1}^{1}\hspace{6cm}\right.\]\vspace{-0.7cm}
\[\hspace{5cm}\left.+\frac{\beta_{1}\beta_{3}M_{1}}{\beta_{2}\theta_{c}^{^{1/2}}}\left(\frac{\delta^{2}\Lambda_{1}^{2}
(\theta_{c},\delta)}{\left(\exp\left(\frac{a_{7}}{u_{2}^{+}}+a_{8}\right)\right)^{2}}
+\frac{a_{5}}{|1+\epsilon(1-\delta)|}\right)\right],\]

\[C_{2}^{2}=\biggl(\frac{b_{1}+|b_{2}|u_{2}^{-}+b_{3}(u_{2}^{+})^{2}}
{\beta_{2}}\biggr)C_{2}^{1}\hspace{6cm}\]\vspace{-0.7cm}
\[\hspace{3.5cm}+\frac{\beta_{1}M_{1}(|b_{2}|+2b_{3}u_{2}^{+})}
{\beta_{2}\theta_{c}^{^{1/2}}}\left(\frac{\delta^{2}\Lambda_{1}^{2}(\theta_{c},\delta)}
{\left(\exp\left(\frac{a_{7}}{u_{2}^{+}}+a_{8}\right)\right)^{2}}
+\frac{a_{5}}{|1+\epsilon(1-\delta)|}\right),\]where
\begin{eqnarray}
\Lambda_{1}(\delta)&=&\frac{\theta_{c}-\delta+1}{\delta^{2}};\nonumber\\[-1.5ex]
\nonumber\\[-1.5ex]
\nonumber\\[-1.5ex]
\Lambda_{2}(\delta)&=&\frac{4\theta_{c}\delta^{^{1/2}}(1+\theta_{c})+\delta(\theta_{c}+1)(\theta_{c}
-\delta+1)+2\theta_{c}\delta^{^{3/2}}(\theta_{c}-\delta+1)}{2\theta_{c}\delta^{^{1/2}}};\nonumber\\[-1.5ex]
\nonumber\\[-1.5ex]
\Lambda_{3}(\delta)&=&\frac{\theta_{c}+1}{2\theta_{c}\delta^{^{1/2}}}.\nonumber
\end{eqnarray}
$\blacksquare$

The following corollary is a consequence of the above Proposition.
\begin{corollary}\label{coro1}
Each scalar field $f_{i}$, $i=1,2$, is a function of class $\mathcal{C}^{1}(\Omega\cup\Gamma,\R)$.
\end{corollary}
By this Corollary and classic results of continuous and differentiable functions
(see \cite{ii}), the following proposition is automatic.
\begin{proposition}\label{propo2}
The scalar field $f_{i}:\Omega\cup\Gamma\rightarrow\R$, $i=1,2$,
satisfies a global Lipschitz condition on $\Omega\cup\Gamma$.
\end{proposition}

Proposition \ref{propo2} guarantees that (\ref{eq9.1}) is a well-posed problem. Therefore,
we finalize this section formulating the following result.
\begin{proposition}\label{propo3}
The problem (\ref{eq9.1}) has a unique
solution $\mathbf{u}\in\mathcal{C}^{1}([0,+\infty),\Omega\cup\Gamma)$ that verifies the given
initial conditions $\mathbf{u}^{0}(t_{0})\in\Omega\cup\Gamma$ for all $t_{0}\in
[0,+\infty)$.
\end{proposition}

\section{Behavior of the solutions of dynamic state for long
times}\label{section4}

Great part of the dynamic analysis of the reactive systems centers
its attention in predicting what will happen to the state variables
when these evolve from an initial time $t_{0}\geq 0$ or equivalent,
to establish the dynamic behavior of the system for all times
$t>t_{0}$. Particularly it is of interest for the engineer to know
how the system will behave when time becomes large, because this
will allow him to determine operation's ranks by means of automatic
control systems designed and implemented consistently with the desired
state. Typically, the state desired at the industrial level corresponds
with a steady-state, therefore, from the mathematical point of view
we are concerned to find out solutions $\mathbf{u}(t)$ of the model
(\ref{eq9.1}) that start at an initial value $\mathbf{u}^{0}$ near
or distant a steady-state $\mathbf{u}^{e}\in\Gamma$, will
tend to this steady-state or another when time $t$ tends to
infinite. For this reason, in this work we combine our analysis with some
reported numerical simulations in \cite{ctacv} that were generated
by means of a computer code based on the method of Runge-Kutta to
fourth order.

\begin{figure}[!htb]
\centering
 \includegraphics[width=6.8cm, height=4.8cm,keepaspectratio]{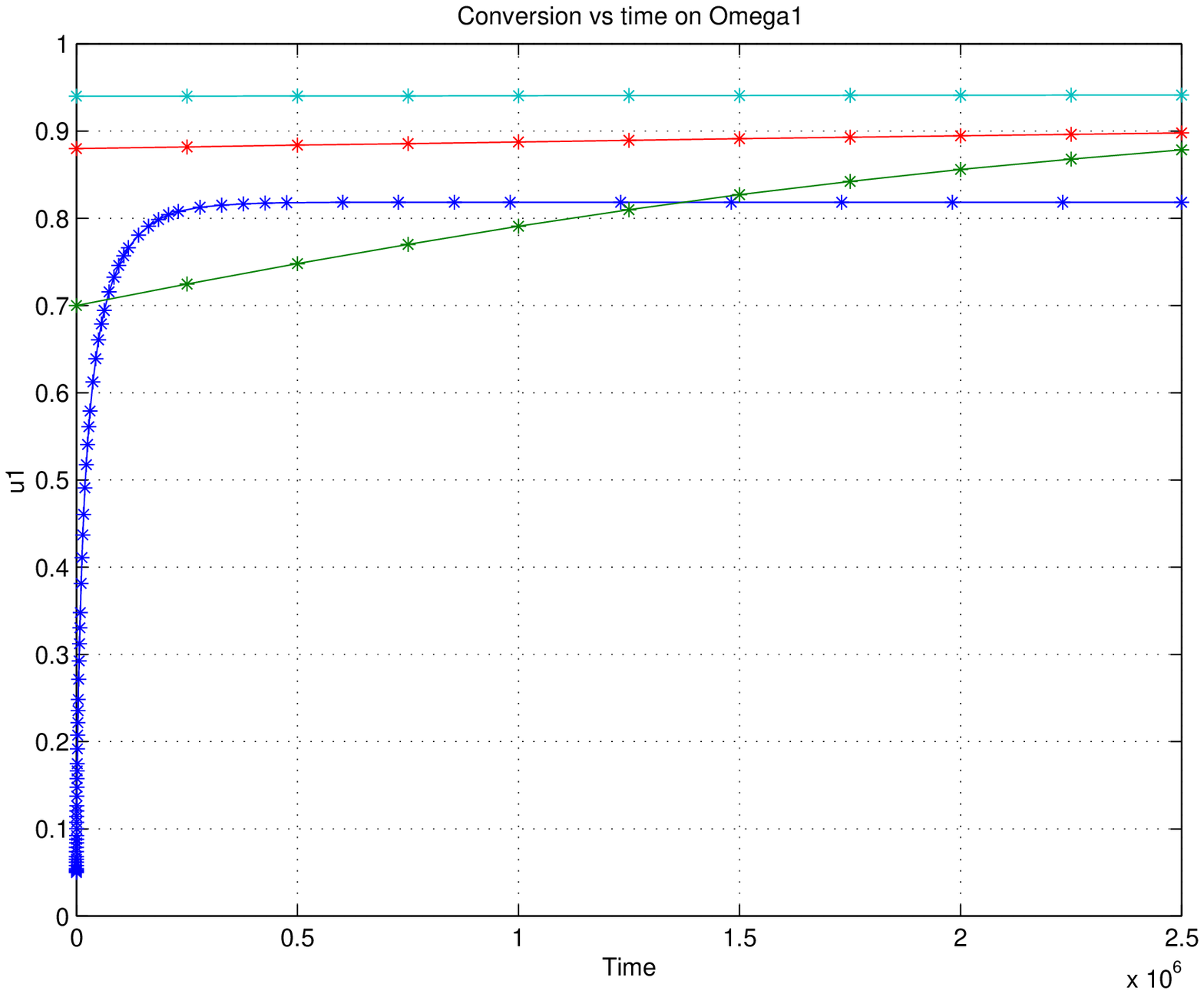}
  \includegraphics[width=6.8cm, height=4.8cm,keepaspectratio]{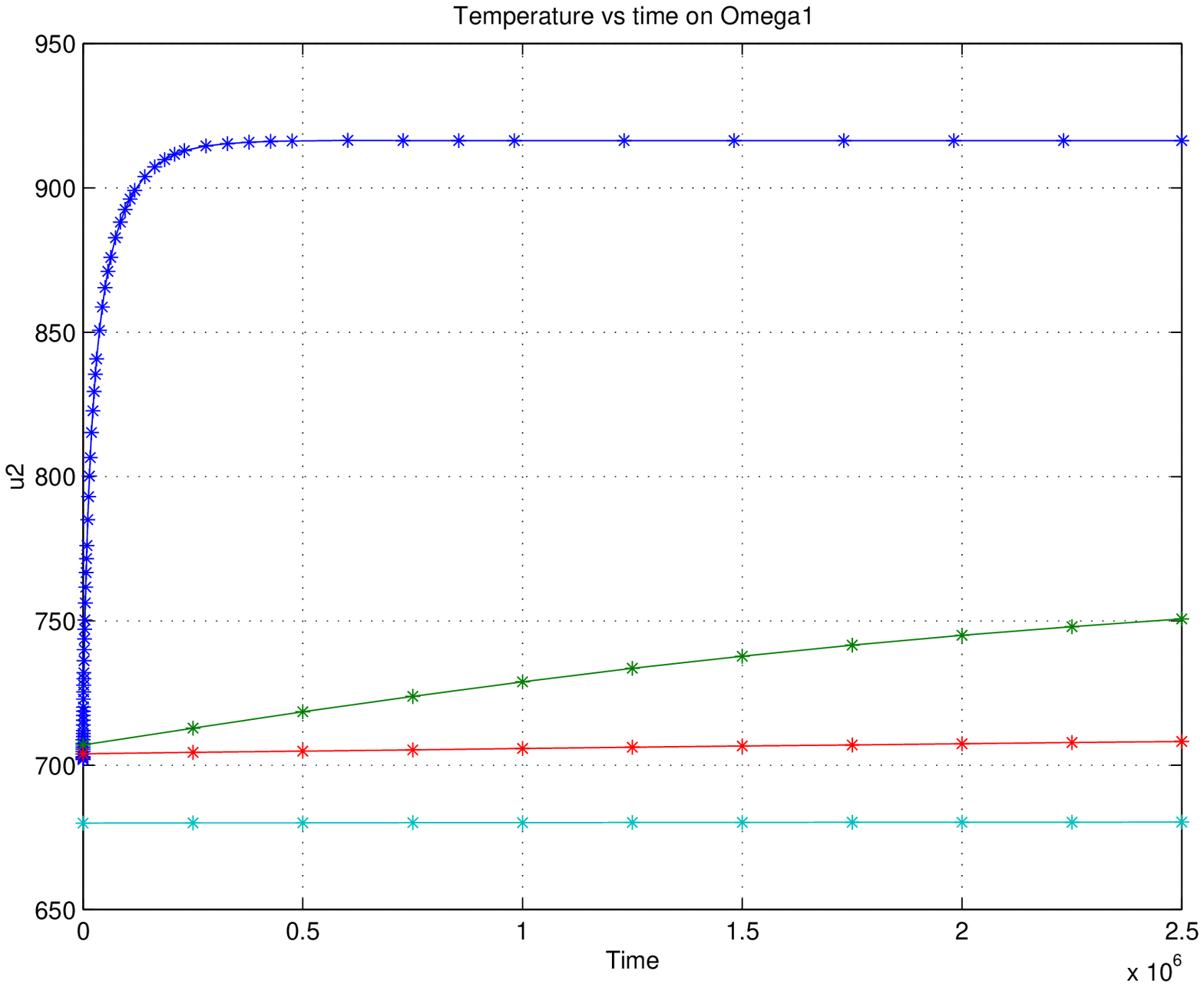}\\
  \caption{\small{Simulations of the conversion and temperature for three initials conditions in $\Omega_{1}$}}\label{simulation1}
\end{figure}

\begin{figure}[!htb]
\centering
\includegraphics[width=6.8cm, height=4.8cm,keepaspectratio]{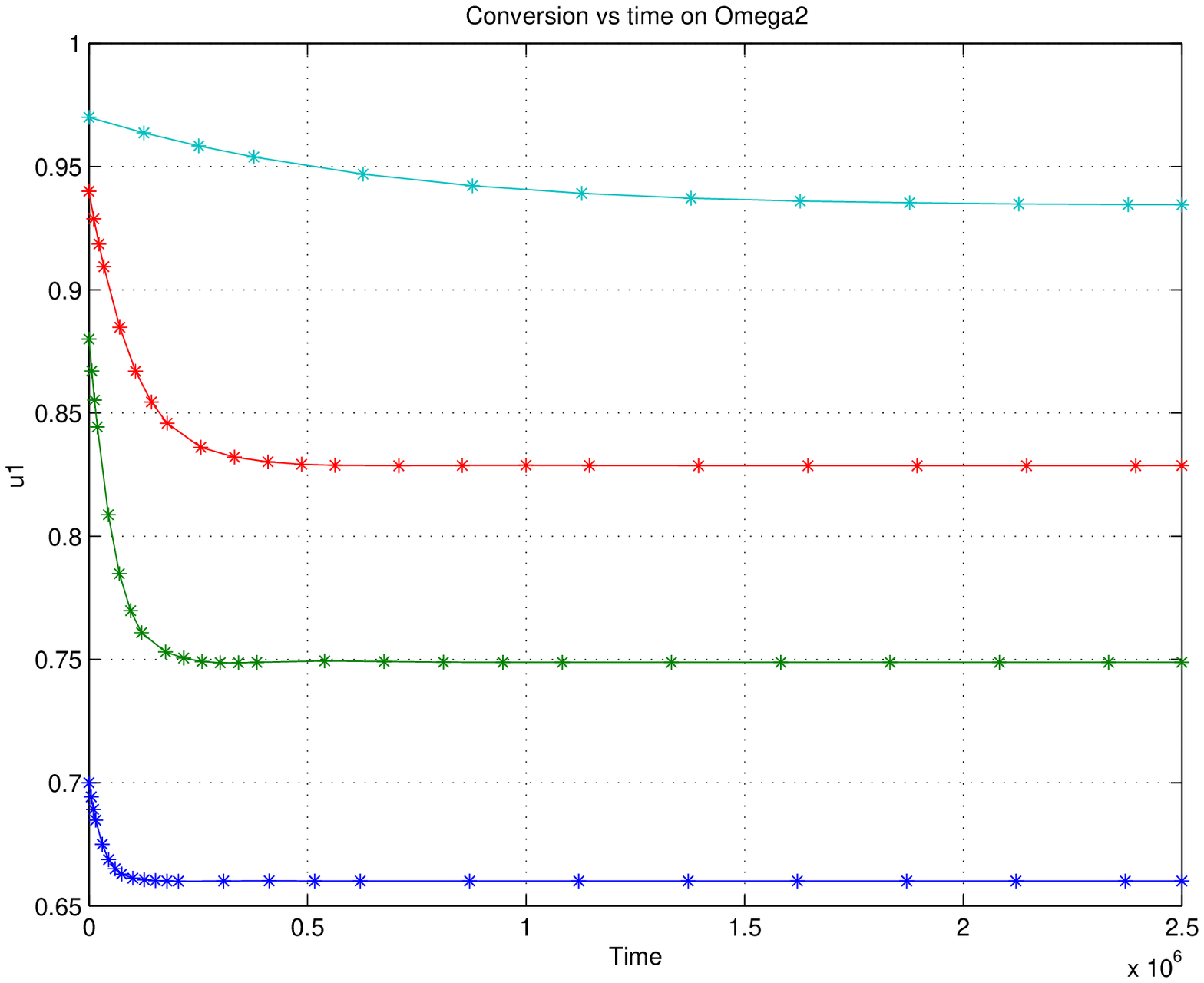}
\includegraphics[width=6.8cm, height=4.8cm,keepaspectratio]{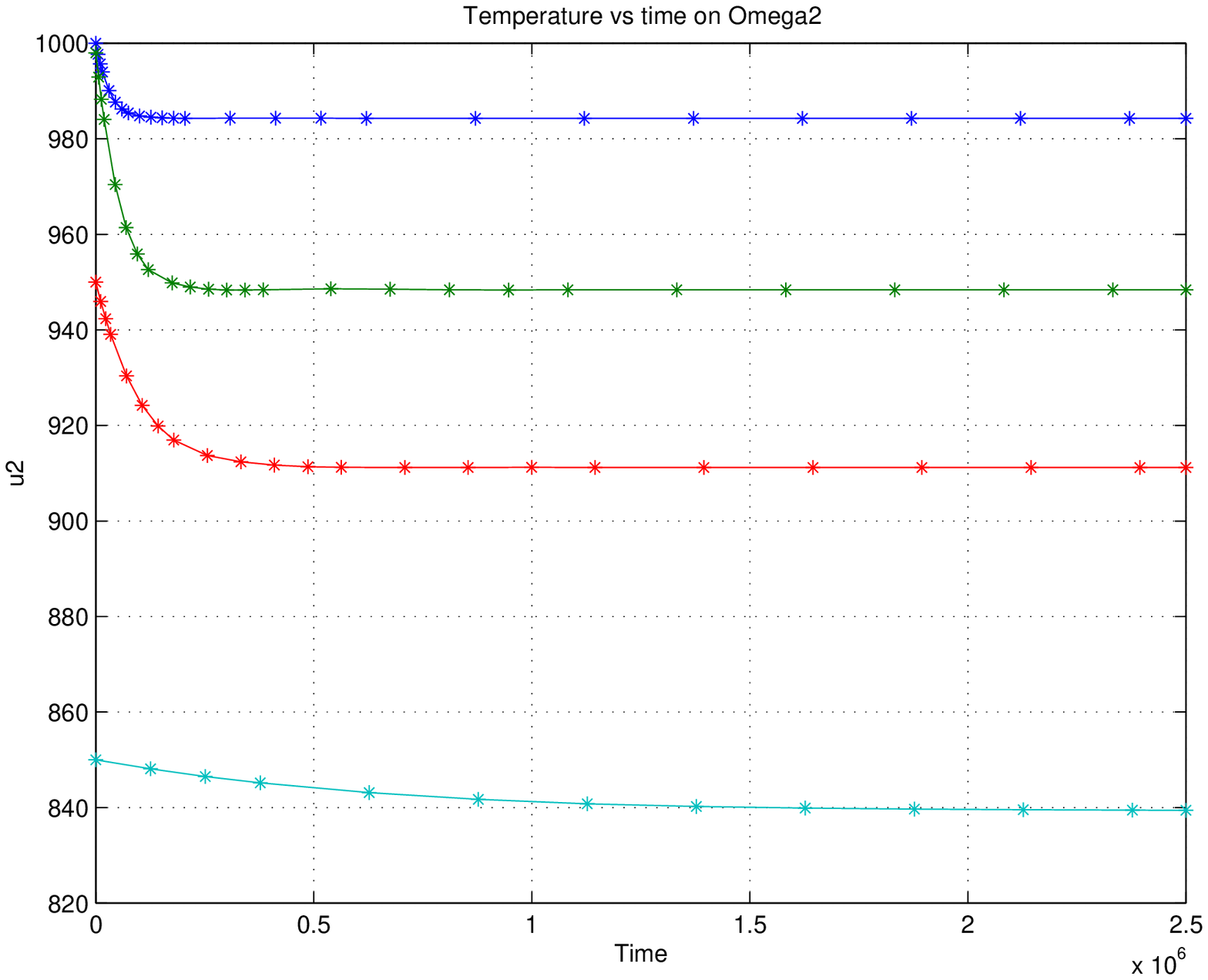}\\
\caption{\small{Simulations of the conversion and temperature for
three initials conditions in $\Omega_{2}$}}\label{simulation2}
\end{figure}

\begin{figure}[!htb]
\centering
\includegraphics[width=7.8cm, height=5.8cm,keepaspectratio]{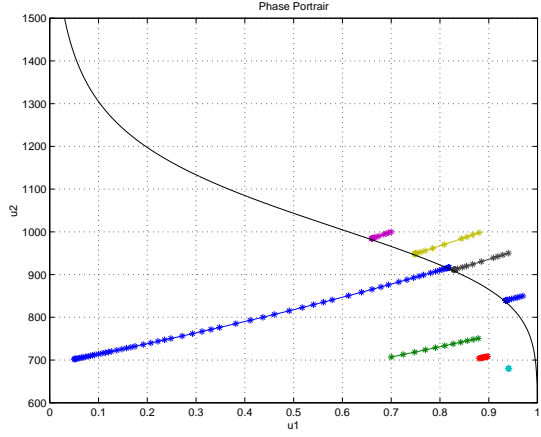}
\caption{Phase portrait}\label{fase1}
\end{figure}

Figures \ref{simulation1} and \ref{simulation2} are
the simulations of four solutions $u_{i}=u_{i}(t)$, $i=1,2$, that
start from four different initial states $u_{i}^{0}=u_{i}(0)$,
$i=1,2$, and Figure \ref{fase1} illustrates a phase portrait where the evolution of the dynamic
states defined by the pair $\mathbf{u}(t)=(u_{1}(t),u_{2}(t))\in\bar{\Omega}$, for all time
$t>t_{0}$, can be demonstrated.

The numerical evidences suggest that there exists sufficient conditions to
state that all $\mathbf{u}=\mathbf{u}(t)$ of the system with
initial value $\mathbf{u}^{0}=\mathbf{u}(0)\in\Omega_{i}$, $i=1$ or
$i=2$, be it sufficiently close or distant to a
steady-state $\mathbf{u}^{e}\in\Gamma$, remain confined on
$\Omega_{i}$, for $i=1$ or $i=2$, and tend to this steady-state when time $t$ tends to infinite. Then
if $\mathbf{u}^{0}=\mathbf{u}(0)\in\Omega_{1}$ and
$\mathbf{\mathbf{u}}(t)\rightarrow\mathbf{u}^{e}$ when
$t\rightarrow\infty$ we can assume that on $\Omega_{1}$ each
component $u_{i}(t)$, for $i=1,2$, is an monotone increasing
function or in fact that for any $\mathbf{u}(t)\in\Omega$ the
vectorial field $\mathbf{u}'(t)=\mathbf{f}(\mathbf{u}(t))$ has
strictly positive sign. Similarly, if
$\mathbf{u}^{0}=\mathbf{u}(0)\in\Omega_{2}$ and
$\mathbf{\mathbf{u}}(t)\rightarrow\mathbf{u}^{e}$ when
$t\rightarrow\infty$ on $\Omega_{2}$ each component $u_{i}(t)$, for
$i=1,2$, is an monotone decreasing function or in fact that for any
$\mathbf{u}(t)\in\Omega$ the vectorial field
$\mathbf{u}'(t)=\mathbf{f}(\mathbf{u}(t))$ has strictly negative
sign. In order to establish these affirmations formally, we have the
following results.
\begin{proposition}\label{propo4}
Let $(\xi_{1},\xi_{2})$ be a point in $\Omega_{1}$ where a solution
of problem (\ref{eq9.1}) begins. The vectorial field
$\mathbf{u}'(t)=\mathbf{f}(\mathbf{u}(t))$ is strictly positive on
$\Omega_{1}$ for all time $t$ in the interval
$I(\xi_{1},\xi_{2})\cap [0,\infty)$.
\end{proposition}

\begproof
Let $(\xi_{1},\xi_{2})\in\Omega_{1}$ be the corresponding point to the initial condition for
problem (\ref{eq9.1}) in $t_{0}=0$, such that
$\mathbf{u}'(0)=\mathbf{f}(\xi_{1},\xi_{2})=\left(f_{1}(\xi_{1},\xi_{2}),f_{2}(\xi_{1},\xi_{2}))\right)$.
Then the sign of vectorial field $\mathbf{f}$ on $\Omega_{1}$ depends on the sign that takes
each component $f_{i}$, $i=1,2$ for all initial point $(\xi_{1},\xi_{2})\in\Omega_{1}$ and
all $t\in I(\xi_{1},\xi_{2})\cap
[0,\infty)$. Indeed
$$f_{1}(\xi_{1},\xi_{2})=\beta_{1}g_{1}(\xi_{1})K_{1}(\xi_{2})K_{2}(\xi_{1})K_{3}(\xi_{1},\xi_{2})
\hspace{0.3cm}\mbox{and}\hspace{0.3cm}f_{2}=f_{1}(\xi_{1},\xi_{2})g_{2}(\xi_{1},\xi_{2}),$$
where if the constant $\beta_{1}$ is positive and by definition, for
all points of the domain, $\Omega$, functions $g_{1}$, $K_{1}$,
$K_{2}$ and $g_{2}$ are strictly positive, then the sign of
$f_{1}(\xi_{1},\xi_{2})$ and $f_{2}(\xi_{1},\xi_{2})$ will be
 positive only if $K_{3}(\xi_{1},\xi_{2})$ has strictly positive sign. This can be verified if we assume
 that there exists a point $(\xi_{1},\xi_{2})\in\Omega_{1}$ for which $K_{3}(\xi_{1},\xi_{2})\leq
 0$. Then if $K_{3}(\xi_{1},\xi_{2})=0$ implies that $\xi_{2}=u_{2}^{e}$, which is a
 contradiction because by definition of subset $\Omega_{1}$ must satisfy that $\xi_{2}<u_{2}^{e}$. Similarly,
 if $K_{3}(\xi_{1},\xi_{2})<0$ then, fixing $\xi_{1}\in
 (0,u_{1}^{\ast})$, we find that $\frac{1}{u_{2}^{e}}>\frac{1}{\xi_{2}}$ implying that $\xi_{2}>u_{2}^{e}$;
 again a contradiction. With this we verified that $K_{3}(\xi_{1},\xi_{2})>0$ for
 all $(\xi_{1},\xi_{2})\in\Omega_{1}$ and we concluded the proof.
$\blacksquare$

The next proposition is proved in a similar way.
\begin{proposition}\label{propo5}
Let $(\zeta_{1},\zeta_{2})$ be a point in $\Omega_{2}$ where a
solution of problem (\ref{eq9.1}) begins. The vectorial field
$\mathbf{u}'(t)=\mathbf{f}(\mathbf{u}(t))$ is strictly negative on
$\Omega_{2}$ for all time $t$ in the interval
$I(\zeta_{1},\zeta_{2})\cap [0,\infty)$.
\end{proposition}

Based on the two previous propositions, with the following corollary we establish the important
property on the behavior of the solutions in $\Omega_{1}$ and $\Omega_{2}$ that was demonstrated in the
numerical simulations.
\begin{corollary}\label{coro2}
For each $i=1,2$, solutions $u_{i}(t)$ of the problem (\ref{eq9.1})
are increasing functions on $\Omega_{1}$ and decreasing functions on
$\Omega_{2}$ for all $t\geq t_{0}$.
\end{corollary}

Finally, the confinement property of the solutions in each $\Omega_{i}$, $i=1,2$, and the tendency of these to a
stationary point of $\Gamma$ when time $t$ tends to infinite is established with the following lemma.
\begin{lemma}\label{lema3}
All solution $\mathbf{u}(t)$ of (\ref{eq9.1}) that begins in the
region defined by subdomain $\Omega_{i}$, for each $i=1,2$, in the
time $t=t_{0}$ remains in that region for all future time $t\geq
t_{0}$, and finally tends to the stationary solution in $\Gamma$.
\end{lemma}
\begproof
The proof shall be presented schematically for $\Omega_{1}$; for
$\Omega_{2}$ is similar.

We assume that a solution
$\mathbf{u}(t)=\left(u_{1}(t),u_{2}(t)\right)$ of (\ref{eq9.1})
leaves the region defined by $\Omega_{1}$ in time $t=t^{\ast}$. Then
$\mathbf{u}'(t^{\ast})=\mathbf{0}$, since the unique way in which a
solution can leave the region defined by $\Omega_{1}$ is crossing
curve $\Gamma$.

Corollary \ref{coro2} assures this, because it indicates that in
that region $\mathbf{u}'(t)>0$ for all $t$. On the other hand the
study of existence and uniqueness of solution of the problem
(\ref{eq9.1}) guarantees that two solutions cannot be cut. For such
reason, as $\Gamma$ is the set of trivial solutions of the problem,
any solution that begins in $\Omega_{1}$ will not cross a solution
in $\Gamma$. Then, this contradicts the fact that
$\mathbf{u}'(t^{\ast})=\mathbf{0}$, assuring that the solutions
initiated in $\Omega_{1}$ remain in that region for all future time
$t\geq t_{0}$. In addition, this implies that each $u_{i}(t)$,
$i=1,2$, is an monotone increasing function of the time for $t\geq
t_{0}$ that is bounded in such region, therefore has a limit when
$t$ tends to infinite. We only must verify that this limit is a
component of all point in $\Gamma$. Indeed, if we denote $\eta_{i}$ the limit of each function $u_{i}(t)$ when $t$ tends
to infinite, then this will imply that $\left|u_{i}(t_{1}) -
u_{i}(t_{2}) \right|$ tends to zero when $t_{1}$ and $t_{2}$ tend to
infinite, since
\begin{eqnarray}
\left|u_{i}(t_{1})-u_{i}(t_{2})\right|&=&
\left|\left(u_{i}\left(t_{1}\right)-\eta_{i}\right)+\left(\eta_{i}-u_{i}\left(t_{2}\right)\right)\right|
\nonumber\\[-1.5ex]
\nonumber\\[-1.5ex]
&\leq&
\left|u_{i}\left(t_{1}\right)-\eta_{i}\right|+\left|u_{i}\left(t_{2}\right)-\eta_{i}\right|.\nonumber
\end{eqnarray}
In particular, let $t_{1}=t$ and $t_{2}=t_{1}+\kappa$ for some
fixed positive number $\kappa$. Then,
$\left|u_{i}(t+\kappa)-u_{i}(t)\right|$ tends to zero when $t$
tends to infinite.
But$$u_{i}(t+\kappa)-u_{i}(t)=\kappa\frac{du_{i}(\tau)}{dt} = \kappa
f_{i}\left(u_{1}(\tau),u_{2}(\tau)\right)$$ where $\tau$ is an arbitrary
number between $t$ and $t+\kappa$. We observe finally that
$f_{i}\left(u_{1}(\tau),u_{2}(\tau)\right)$ must tend to
$f_{i}\left(\eta_{1},\eta_{2}\right)$ when $t$ tends to infinite.
Therefore, $f_{i}\left(\eta_{1},\eta_{2}\right)=0$ for each $i=1,2$,
and with this we conclude the proof.
$\blacksquare$

\begin{figure}[!htb]
\centering
\includegraphics[width=6.5cm, height=4.5cm,keepaspectratio]{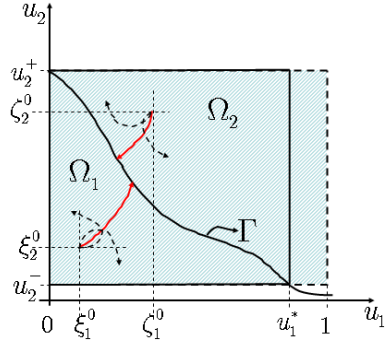}
\caption{Situations that cannot happen with
$\mathbf{u}(t)$}\label{domain2}
\end{figure}

Lemma \ref{lema3} guarantees that solutions of the system that begin
either in subdomains, $\Omega_{1}$ and $\Omega_{2}$, tend to a
steady-state located on curve $\Gamma$ when the time tends to
infinite whatever the initial starting condition within each
region is, therefore situation shown in the figure \ref{domain2}, and
illustrated on each subdomain for the curves drawn up by segments
that start in $(\xi_{1}^{0},\xi_{2}^{0})\in\Omega_{1}$ and in
$(\zeta_{1}^{0},\zeta_{2}^{0})\in\Omega_{2}$, cannot happen.

\section{Brief discussion}\label{section5}

The two regions $\Omega_{1}$ and $\Omega_{2}$
characterized in this work, are the regions in which both direct and inverse chemical reaction
take place. This can be affirmed because we demonstrated with Corollary (\ref {coro2}) that in $\Omega_{1}$ when
increasing the conversion of $\mbox{SO}_{2}$, the temperature of the system
increases too; indeed, $\Omega_{1}$ is the region where the
exothermic character of the reaction predominates. Similarly, it is
proved that the region defined by $\Omega_{2}$ is where the
endothermic character of the reaction predominates (diminution of
the conversion of $\mbox{SO}_{2}$ and temperature of the system).

On the other hand, with Lemma \ref{lema3} we demonstrated that the
solutions of the model (\ref{eq9.1}) tend to a steady-state whatever the starting state in $\Omega_{i}$, for
each $i=1,2$, and remain confined in this region. This behavior  is consistent with the physical phenomenon, since, experimentally we know that with very long time steps the
reaction tends to the equilibrium because always there exist infinitesimal changes
 in the conversion and temperature; which define
states called quasi-steady states.

Also we know that the course of
reaction changes if the system is perturbed providing or removing
energy from it. This perturbation will locate the reaction in a new initial
state on the same subdomain or on the other subdomain; in
the latter, the change of subdomain is not a natural behavior
of the system. This change is caused by an external factor that modifies the
initial conditions of the problem; therefore, the solutions of the
mathematical model must remain in the origin region as long as they
are not perturbed.

\section{Conclusions}\label{section6}

The solutions of a mathematical model that is used to analyze the
dynamic behavior of reversible reaction
$\mbox{SO}_{2\,(\mathrm{g})}\,+\,
\frac{1}{2}\,\mbox{O}_{2\,(\mathrm{g})}\,\rightleftharpoons \,
\mbox{SO}_{3\,(\mathrm{g})}$ carried out in a catalytic reactor,
were studied qualitatively by means of the abstraction of the model
in terms of the state variables: conversion of the $\mbox{SO}_{2}$
and temperature of the system. In this sense, we demonstrated that
the model is a well-posed Cauchy problem; i.e., there exists an
unique solution for each initial condition related to the state
variables.

The trivial solutions of the mathematical model correspond with the
steady-states of the reactive system and conform a continuous front on the phase portrait for conversion versus temperature. In
fact, the phase portrait was divided in two separated regions by the
continuous steady-states, and we demonstrated that in a region
the reaction advances exothermically and in the
other region it advances endothermically as we expect from the physicochemical point of view. Also, we proved
that when time becomes sufficiently large, conversion of the
$\mbox{SO}_{2}$ and temperature of the reactive system will remain
near some steady-state whatever the point in the phase portrait from
which the state variables begin. All these theoretical results were
complemented with numerical simulations by means of which we
observed a priori the hypotheses and conclusions of the
Propositions, Corollaries and Lemmas that we presented in this
article.

  %--------------------------------------------------------------------------------------
%\section*{ACNOWLEDGMENTS}
%--------------------------------------------------------------------------------------

% \newpage

%----------------------------------------- ---------------------------------------------

\end{document}